\documentclass[10pt,epsf]{article}

\usepackage{graphicx}
\usepackage{indentfirst}
\usepackage{color}

\begin{document}

\title{\bf{A Particle Probing Thermodynamics in Three-Dimensional Black Hole}}

\date{}
\maketitle

\begin{center}\author{Bogeun Gwak}\footnote{rasenis@sogang.ac.kr} and \author{Bum-Hoon Lee}\footnote{bhl@sogang.ac.kr}\\ \vskip 0.25in $^{1,2}$\it{Department of Physics and Center of Quantum Spacetime, Sogang University, Seoul 121-742, Korea} \end{center} \vskip 0.6in

{\abstract
{We have shown the thermodynamic changes in a 2+1-dimensional rotating black hole when it absorbs a particle. The microscopic changes which the black hole undergoes are interpreted using the AdS/CFT correspondence. Using particle absorption phenomena, we formulate an irreducible mass and black hole entropy directly related to the particle momenta. We describe the black hole evolution that preserves the 2nd law of thermodynamics. The numbers of microstates which evaluate the entropy are analyzed at each of the black hole evolution stages.}}

\thispagestyle{empty}
\newpage
\setcounter{page}{1}
\section{Introduction}
The information regarding the black hole mass and momenta can be obtained from its horizon surface area as a Bekenstein-Hawking entropy\,\cite{BH1}. The black hole energy states are described on the surface without any relation to the internal structure. It also satisfies the 2nd law of thermodynamics for a particle absorption\,\cite{BGwak}. The particle or matter falling into a black hole can increase or decrease mass or angular momentum, but always increase the entropy. The microscopic origin of the entropy is explained from a quantum field theory\,\cite{Org}. Especially, using D-brane description, the result for the Bekenstein-hawking entropy is reproduced as the logarithm of the microstate counting of the D-brane configurations in the specific black hole\,\cite{entD}. Using AdS/CFT dictionary, some AdS black hole entropy is interpreted as the asymptotic growth of states described by a conformal field theory(CFT)\,\cite{Strom1,acc1}\,. Recently, for asymptotically flat spacetime, the Bekenstein-Hawking entropy for the Kerr black hole has been interpreted as a microscopic entropy for the specific CFT, which leads to the conjecture of Kerr/CFT correspondence\,\cite{kerrCFT}. The earlier works regarding a rotating black hole entropy are mainly about Banados-Teitelboim-Zanelli (BTZ) black hole\,\cite{BTZ1,BTZ2}. It is a rotating black hole solution with a cosmological constant in 2+1 dimensions. A BTZ black hole gives enough evidences to understand rotating black hole properties algebraically. Some black holes based on string theory have the similar properties of the BTZ black hole, so obtaining the entropy and the grey-body factor for the BTZ black hole can give direct these black holes\,\cite{Crp}. In AdS/CFT duality, these properties are interpreted as 2-dimensional CFT. 

In this paper, we microscopically interpret the 2+1-dimensional rotating black hole's thermodynamic changes in a particle absorption using AdS/CFT correspondence. The BTZ black hole entropy has been derived by Strominger\,\cite{Strom1}. In his derivation, the AdS black hole entropy is obtained by Cardy's formula\,\cite{Cardy} using diffeomorphism operators. We shortly give a review about this work in the first section. Using this description, the microstate changes which the black hole undergoes when a particle falls into it is shown. It is the importance of the particle absorption that preserves the 2nd law of thermodynamics\,\cite{BGwak}. The black hole mass is simply observed as the rest mass plus rotational energy, so artificial controls about the black hole mass and angular momentum break the thermodynamic laws and cannot show any physical evolutions. However, if we control black hole properties by the mass and momenta carried by a particle falling into the black hole, the black hole properties changes substantially conserve the thermodynamic laws. Through the particle absorption applied to the black hole evolutions, the changes of the black hole configuration can be traced in the real universe. From the CFT point of view, the particle energy and momentum changes the black hole energy states which are composed as sum of quantum number for the left and right moving operators, and the particle effects work differently for each ratios of the black hole mass and angular momentum. We have shown what a 2+1-dimensional rotating black hole undergoes and how its microstates change during a particle absorption.

The paper is organized as follows. First, we briefly review the derivation of the Bekenstein-Hawking entropy formula as the asymptotic growth of states described by a CFT. Secondly, thermodynamic changes in a black hole are obtained in terms of the infalling particle momenta. Thirdly, the microstate excitations on the CFT side are interpreted as a particle carrying momenta. Fourthly, thermodynamic and microscopic changes are analyzed at each of the black hole evolution stages. Finally, we summarize and discuss our results and prospects.

\section{Three-Dimensional Black Hole and Conformal Field Theory}
In this section, we briefly review about the three-dimensional rotating black hole entropy form the asymptotic growth of states described by a conformal field theory(CFT)\,\cite{Strom1}\,. The gravitational action in three-dimensional spacetime coupled to matter is given by,
\begin{eqnarray}
S=\frac{1}{16\pi G}\int d^3 x \sqrt{-g}(R+2\ell^{-2}) + S_{matter} + S_{surface}\,,
\end{eqnarray}
where $S_{matter}$ and $S_{surface}$ are the matter action and the surface term, not to be considered in our paper. The cosmological constant $\Lambda$ is $-\ell^{-2}$\,. Note that our discussion is valid in the semiclassical regime, $\ell\gg G$\,. The AdS$_3$ vacuum solution obtained from the action is,
\begin{eqnarray}
\label{ads3}
ds^2=-(\ell^{-2} r^2+1)dt^2+(\ell^{-2} r^2+1)^{-1}dr^2+r^2d\phi^2\,.
\end{eqnarray}
The AdS$_3$ part is included in the $SL(2,R)$ group manifold, and the isometry group of that is $SL(2,R)_L\otimes SL(2,R)_R\,$. The boundary condition should be determined to define the full quantum theory on AdS$_3$. Thus, the condition required is written as,
\begin{eqnarray}
\label{bndc1}
&&g_{tt}=-\ell^{-2} r^2+\mathcal{O}(1)\,,\,\,g_{t\phi}=\mathcal{O}(1)\,,\,\,g_{tr}=\mathcal{O}(1/r^3)\,,\\
&&g_{rr}=\ell^{2} r^{-2}+\mathcal{O}(1/r^4)\,,\,\,g_{r\phi}=\mathcal{O}(1/r^3)\,,\,\,g_{\phi\phi}=r^2+\mathcal{O}(1)\,.\nonumber
\end{eqnarray}
The non-trivial diffeomorphism is obtained from the vector fields preserving the boundary condition. The generators of the diffeomorphism are defined as $L_n$ and $\bar{L}_n$ in $-\infty<n<\infty\,$. The generators obey the Virasoro algebra,
\begin{eqnarray}
&&[L_m,L_n]=(m-n)L_{m+n}+\frac{c}{12}(m^3-m)\delta_{m+n,0}\,,\\
&&[\bar{L}_m,\bar{L}_n]=(m-n)L_{m+n}+\frac{c}{12}(m^3-m)\delta_{m+n,0}\,,\,\,[L_m,\bar{L}_n]=0\,,\nonumber
\end{eqnarray}
where the central charge $c=3\ell/2G\,$. Therefore, quantum gravity on AdS$_3$ is a conformal field theory with a central charge $c=3\ell/2G\,$\cite{BrHn1}. The CFT lies on the $(t,\phi)$ cylinder at spatial infinity.

The BTZ black hole\,\cite{BTZ1,BTZ2} which preserves the required boundary condition in eq.~(\ref{bndc1}) can be obtained by a discrete identification of eq.~(\ref{ads3}) and written\,\cite{Strom1} as,
\begin{eqnarray}
\label{metric1}
&&ds^2=-N^2dt^2+\rho^2(N^{\phi}dt+d\phi)^2+\frac{r^2}{N^2 \rho^2}dr^2\,,\\
&&N^2=\frac{r^2(r^2-r_h^2)}{\ell^2\rho^2}\,,\,\,N^\phi=-\frac{4GJ}{\rho^2}\,,\,\,\rho^2=r^2+4GM\ell^2-\frac{1}{2}r_h^2\,,\,\,r_h^2=8G\ell\sqrt{M^2\ell^2-J^2}\,,\nonumber
\end{eqnarray}
where $M$ and $J$ are the black hole mass and angular momentum. Note that the black hole is in the Hilbert space of the CFT and goes to eq.~(\ref{ads3}) when $M=-\frac{1}{8G}\,$. The black hole mass and angular momentum are redefined in terms of additional constants $L_0$ and $\bar{L}_0$ which will be matched with the generators of the previous CFT. The redefined mass and angular momentum is,
\begin{eqnarray}
\label{mj1}
M=\frac{1}{\ell}(L_0+\bar{L}_0)\,,\,\,J=L_0-\bar{L}_0\,.
\end{eqnarray}
The black hole solution is assumed to be the excitations of the vacuum obtained as $M=J=0\,$. The vacuum is,
\begin{eqnarray}
\label{zerobh1}
ds^2=-\frac{r^2}{\ell^2}dt^2 +r^2d\phi^2+\frac{\ell^2}{r^2}dr^2\,.
\end{eqnarray}
It is a different spacetime, globally, but there exists only one constant curvature metric locally in 3 dimensions. Therefore, it is equivalent to AdS$_3$, locally. Making the black hole locally equivalent to AdS$_3$, the black hole mass is the meaning of AdS$_3$ vacuum. In super conformal field theory(sCFT), the Ramond ground state of the periodic boundary condition has zero mass, which corresponds to the zero mass solution of eq.~(\ref{zerobh1})\,. For anti-periodic condition, the Neveu-Schwarz ground state gets a mass shift by $L_0=\bar{L}_0=-c/24\,$. The central charge of the CFT lying on AdS$_3$ and eq.~(\ref{mj1}) gives $M=-1/8G\,$\,\cite{CoHn}, and further evidences are in \cite{Strom1}. The asymptotic growth of states for given the CFT can be shown to obey Cardy's formula\,\cite{Cardy}\,,
\begin{eqnarray}
S=2\pi\sqrt{\frac{c n_R}{6}}+2\pi\sqrt{\frac{c n_L}{6}}\,,
\end{eqnarray}
where $n_R$ and $n_L$ are the eigenvalues of $L_0$ and $\bar{L_0}\,$ and $n_R+n_L\gg c\,$ in the semiclassical regime. Using eq.~(\ref{mj1}), the asymptotic growth of states is rewritten  in terms of the BTZ black hole's mass and angular momentum.
\begin{eqnarray}
\label{ent3}
S_{BH}=\pi\sqrt{\frac{\ell(\ell M +J)}{2G}}+\pi\sqrt{\frac{\ell(\ell M -J)}{2G}}\,,
\end{eqnarray} 
which is exactly the same as the Bekenstein-Hawking entropy\,\cite{Strom1}.

\section{Thermodynamics in the Particle Absorption Process}
The black hole properties are controlled by the particle momenta. To obtain the first-order geodesic equations, the separation of variable method is applied similar to the case of Kerr black hole\,\cite{Carter}. The Hamiltonian and Hamilton-Jacobi action are,
\begin{eqnarray}
\mathcal{H}=\frac{1}{2}g^{\mu \nu}p_\mu p_\nu\,,\,\,S=\frac{1}{2}m^2\lambda -Et+L\phi +S_r(r)\,,
\end{eqnarray}
where the conserved quantities are obtained from translation symmetries for $t$ and $\phi$ of the metric. The inverse metric is,
\begin{eqnarray}
g^{tt}=-\frac{1}{N^2}\,,\,\,g^{t\phi}=g^{\phi t}=\frac{N^\phi}{N^2}\,,\,\,g^{\phi \phi}=\frac{1}{\rho^2}-\frac{{N^\phi}^2}{N^2}\,,\,\,g^{rr}=\frac{N^2 \rho^2}{r^2}\,.
\end{eqnarray}
The first-order geodesic equations are obtained as follows,,
\begin{eqnarray}
&&\dot{r}=p^r=\frac{N^2\rho^2}{r^2}\sqrt{R(r)}\,,\,\,\dot{t}=\frac{E-N^\phi L}{N^2}\,,\,\,\dot{\phi}=\frac{-{N^\phi}^2L +EN^\phi}{N^2}-\frac{L}{\rho^2}\,,\\
&&R(r)=\frac{r^2}{N^2 \rho^2}\left[-m^2+\frac{1}{N^2}\left[E^2+2N^\phi E L +{N^\phi}^2 L^2\right]-\frac{L^2}{\rho^2}\right]\,\nonumber.
\end{eqnarray}
Using radial geodesic equation, the dispersion relation described by the particle location and momenta is constructed as,
\begin{eqnarray}
E^2+2N^\phi E L +{N^\phi}^2 L^2-\frac{r^2{p^r}^2}{N^2 \rho^2}-m^2-\frac{L^2}{\rho^2}=0\,.
\end{eqnarray}
The particle energy is a positive valued solution, which lies in the future-forwarding geodesics. The particle energy is absorbed to the black hole when it reaches to the outer horizon, $r_h$. The particle energy constructed in terms of momenta at the horizon is obtained as,
\begin{eqnarray}
\label{pe1}
E=\frac{4GJ}{\rho_h^2}L + \frac{r_h|p^r|}{\rho_h}\geq E_{min}=\frac{4GJ}{\rho_h^2} L\,,\,\,\rho_h=\sqrt{4GM\ell^2+\frac{1}{2}r_h^2}\,,
\end{eqnarray} 
where the energy has a minimum from the positive sign of the coefficient $E$ square. It is achieved when $p^r=0$ at the horizon. The black hole mass and angular momentum are changed as much as those of the particle, so it can be written as $\delta M = E$ and $\delta J = L$\,. The rewritten eq.~(\ref{pe1}) is given by,
\begin{eqnarray}
\label{bhe1}
\delta M=\frac{4GJ}{\rho_h^2}\delta J + \frac{r_h|p^r|}{\rho_h}\geq E_{min}\,.
\end{eqnarray}
The black hole mass is a sum of the irreducible mass and rotational energy\,\cite{ChrisRu, Smarr}. After removing the rotational energy, the irreducible mass and its change are obtained,
\begin{eqnarray}
\label{mir1}
M_{ir}=({\frac{1}{2}r_h^2+4GM\ell^2})^{\frac{1}{4}}\,,\,\,\delta M_{ir}=\frac{|p^r|r_h\ell}{4\sqrt{\rho_h(M^2\ell^2-J^2)}}\,,
\end{eqnarray}
where the change in the irreducible mass depends only on the particle radial momentum. The Bekenstein-Hawking entropy $S_{BH}$ can be rewritten in terms of the irreducible mass,
\begin{eqnarray}
\label{ent1}
S_{BH}=\frac{\mathcal{A}}{4G}=\frac{2\pi M_{ir}^2}{4G}=\frac{\pi \sqrt{16GM\ell^2+2r_h^2}}{4G}\,.
\end{eqnarray}
The change of the black hole entropy after the particle absorption is,
\begin{eqnarray}
\label{delent1}
\delta S_{BH}=\frac{\pi M_{ir}\delta M_{ir}}{G}=\frac{\pi |p^r| r_h \ell}{4G\sqrt{M^2 \ell^2 -J^2}}\,,
\end{eqnarray}
which depends only on the particle radial momentum. The changes in the thermodynamic properties  of the black hole can be classified into four cases according to the particle rotating direction and radial momentum at the horizon. If the particle rotates in the same direction as that of the black hole, the black hole mass is increased. For a particle with a non-zero radial momentum, the entropy and irreducible mass are increased from eq.~(\ref{mir1}) and (\ref{ent1})\,. It is interpreted as an irreversible process, since $\delta S_{BH}>0\,$. The horizon location is changed due to the absorption, because the black hole mass and angular momentum also change. From the increase in $M_{ir}\,$ from eq.~(\ref{mir1})\,, the size of the horizon is shown to be,
\begin{eqnarray}
\label{rh1}
\delta r_h = \frac{32G^2\ell^4 M}{r_h^2 \rho_h}|p^r|-\frac{16G^2\ell^2 J}{r_h\rho_h^2}L\,.
\end{eqnarray}
If the particle rotates in the same direction as that of the black hole, the horizon becomes bigger than before after the black hole absorbs a particle of a radial momentum bigger than $\frac{r_hJL}{2\ell^2\rho_h M}$\,. For a particle of radial momentum smaller than $\frac{r_hJL}{2\ell^2\rho_h M}$\,, the horizon becomes smaller. The black hole rotational energy increases due to the larger $M_B$ and $J$ as contributed by the particle momenta. For the specific case of a particle with zero radial momentum, the entropy and irreducible mass remain unchanged. In other words, $\delta S_{BH}\sim \delta M_{ir} =0\,$. It can be interpreted as a reversible process and the black hole horizon becomes always smaller than before in eq.~(\ref{rh1}). The black hole rotational energy increases due to the bigger $M_B$ and $J$ as contributed by the particle momenta.

If the particle rotates in the opposite direction to that of the black hole, the black hole mass can decrease due to removal of the black hole rotational energy. If the particle radial momentum is smaller than $\frac{4GJL}{\rho_h}|L|$\,, the black hole mass gets decreased from eq.~(\ref{bhe1})\,. If the particle radial momentum is larger than $\frac{4GJL}{\rho_h}|L|$\,, the black hole mass is increased. For a particle with a non-zero radial momentum, $p^r\,$, the entropy and irreducible mass still increase in an irreversible process. The horizon becomes bigger than before, for any values of the radial momentum. The black hole rotational energy decreases due to the smaller $J$ removed by a particle momentum. For the specific case of a particle with zero radial momentum, the entropy and irreducible mass remain unchanged. In other words, $\delta S_{BH}\sim \delta M_{ir} =0\,$ in a reversible process. The black hole horizon becomes always bigger than before in eq.~(\ref{rh1}). The black hole rotational energy decreases due to the smaller $M_B$ and $J$ as removed by the particle momenta.

\section{Microstate Excitations in the Dual CFT}
The change of the black hole entropy in eq.~(\ref{delent1}) is equivalent to that of eq.~(\ref{mir1}). Thus, it can be written as the changes of $n_R$ and $n_L$\,,
\begin{eqnarray}
\delta S_{BH} = \delta \left[\,\pi\sqrt{\frac{\ell(\ell M+ J)}{2G}}+\pi\sqrt{\frac{\ell(\ell M - J)}{2G}}\,\right]=\pi\sqrt{\frac{c}{6n_R}}\delta n_R+\pi\sqrt{\frac{c}{6n_L}}\delta n_L\,,
\end{eqnarray}
where the central charge, $c$ is invariant since the boundary is still $AdS_3$ ever after the absorption. The entropy change due to the microstate change should be same as that of Bekenstein-Hawking entropy, so the changes in total and each state numbers satisfying the condition is obtained,
\begin{eqnarray}
\label{deln1}
\delta n_R=\frac{\ell r_h}{2\rho_h}|p^r|+\frac{4G\ell J +\rho^2_h}{2\rho^2_h} L\,,\,\,\delta n_L=\frac{\ell r_h}{2\rho_h}|p^r|+\frac{4G\ell J -\rho^2_h}{2\rho^2_h} L\,,\,\,\delta n_R+\delta n_L=\frac{\ell r_h}{\rho_h}|p^r|+\frac{4G\ell J}{\rho^2_h} L\,.
\end{eqnarray}
Note that $4G\ell J -\rho^2_h\leq 0$ for $\ell M \geq J$\,. Each state numbers behave differently for given particle momenta. The particle radial momentum contributes similarly to $n_R$ and $n_L$, but the angular momentum contributes differently. The change in total number of states is not matched with the entropy change, because the entropy is increased to maximize the state degeneracy. Therefore, the microstate structure and the degeneracy can be elucidated from the state number changes due to a particle absorption.

It is categorized into six cases with the particle momenta. If the black hole absorbs the zero radial momentum particle, the entropy is not changed. How much does the particle rotation energy affect the black hole microstates, can be formed in these cases. For the same rotating direction, $n_R$ increases, but $n_L$ decreases. The total number of states increase as large as $\frac{4G\ell J}{\rho^2_h}L>0$\,. For the opposite direction to the black hole, $n_R$ decreases, but $n_L$ increases. The total number of states decrease, since $\delta n_R + \delta n_L=\frac{4G\ell J}{\rho^2_h}L<0$\,. 

If the black hole absorbs non-zero radial momentum particle, the entropy and $n_R$ increase. When a particle rotates in the same direction as that of the black hole and has a radial momentum $p^r$ bigger than $\frac{\rho^2_h-4G\ell J}{\ell r_h \rho_h}L\,$, $n_L$ increases. For $p^r \leq \frac{\rho^2_h-4G\ell J}{\ell r_h \rho_h}L\,$, $n_L$ decreases, but the total number of states increases. If the black hole absorbs the non-zero radial momentum particle, the entropy and $n_R$ are increased. When a particle rotates in the opposite direction to the black hole and has a radial momentum $p^r$ bigger than $-\frac{\rho^2_h+4G\ell J}{\ell r_h \rho_h}L\,$, $n_R$ increases. For $\frac{4G\ell J}{\rho_h \ell r_h}\leq p^r \leq -\frac{\rho^2_h+4G\ell J}{\ell r_h \rho_h}L\,$, $n_R$ decreases, but the total number of states increases. In addition, if the radial momentum is smaller than $\frac{4G\ell J}{\rho_h \ell r_h}$, $n_R$ decreases, and the total number of states also decreases. 

\section{Phase and Microstate Changes in Particle Absorption}
We consider the black hole evolution by adding a particle. Starting with a black hole given mass and angular momentum, the evolution shows how drastically black hole properties get changed. Also, the changes are interpreted as microstate changes. In the process, the 2nd law of thermodynamics is conserved when we change the black hole properties, so it is more similar to what an observer can watch in 3-dimensional spacetime. Especially, the black hole horizon disappears for the specific case $J=\ell M$. The possibility of this to happen will be shown from the particle absorption point of view. For a black hole with mass $M_1$ and angular momentum $J_1=0$, the horizon and microstates changes depend on a particle radial momentum. When the particle radial momentum is in the range of $0\leq p^r < L/\ell\,$, the effects are, 
\begin{eqnarray}
0\leq p^r < \frac{L}{\ell}\,\rightarrow\, 0\leq\delta r_h <\frac{32G^2\ell^3 M_1}{r_h^2 \rho_h}L\,,\,\,-\frac{L}{2}\leq \delta n_L<0\,,\,\,\frac{L}{2}\leq \delta n_R < L\,,\,\,0\leq\delta n_{tot}<\frac{L}{2}\,,
\end{eqnarray}
where $\delta n_{tot}=\delta n_L+\delta n_R\,$. The entropy and total quantum number increase, but the number of left movers decreases. The reason of decrease in $n_L$ is due to the adding angular momentum. The horizon is always bigger than before. If a particle has a radial momentum bigger than $L/\ell$, all of these quantities increase as,
\begin{eqnarray}
p^r \geq \frac{L}{\ell}\,\rightarrow\, \delta r_h \geq\frac{32G^2\ell^3 M_1}{r_h^2 \rho_h}L\,,\,\,\delta n_L\geq0\,,\,\,\delta n_R \geq L\,,\,\,\delta n_{tot}\geq\frac{L}{2}\,,
\end{eqnarray}
A large radial momentum of the particle increases $n_L$ more than discounting from adding angular momentum in $\delta n_L\,$. The black hole mass gets increased not only by means of the irreducible mass, but also by the rotational energy carried by a particle. After the particle absorption, the black hole has angular momentum $0<J_2<\ell M_2$ and mass $M_2>M_1\,$. The behaviors depend on the particle radial momentum. The phenomena is divided into three different cases. If the particle radial momentum is bigger than $-\frac{4GJ\ell - \rho_h^2}{\ell r_h \rho_h}L\,$, the changes are,
\begin{eqnarray}
|p^r|\geq -\frac{4GJ_2\ell - \rho_h^2}{\ell r_h \rho_h}L\,\rightarrow\,\delta r_h\geq \frac{16G^2 \ell^2}{r_h^3\rho_h^3}(r_h^2(\ell M_2-J_2)+8\ell^3G M_2)\,,\,\,\delta n_L\geq0\,,\,\,\delta n_R\geq L\,,\,\,\delta n_{tot}\geq L\,,
\end{eqnarray}
so the large value of the radial momentum increases $n_L$\,, $n_R$, and $n_{tot}$. Although the negative contribution of angular momentum in $n_L$\,, the radial momentum increases $n_{tot}$. If the particle radial momentum is in between $\frac{Jr_h}{2\ell^2M \rho^2}$ and $-\frac{4GJ\ell - \rho_h^2}{\ell r_h \rho_h}L\,$, the changes are,
\begin{eqnarray}
&&\frac{J_2r_h}{2\ell^2 M_2 \rho_h}L\leq |p^r|<-\frac{4GJ_2\ell - \rho_h^2}{\ell r_h \rho_h}L\,\rightarrow\,0\leq\delta r_h< \frac{16G^2 \ell^2}{r_h^3\rho_h^3}(r_h^2(\ell M_2-J_2)+8\ell^3G M_2)\,,\\
&&-\frac{M_2\ell-J_2}{2\ell M_2}L\leq\delta n_L<0\,,\,\,\frac{M_2\ell+J_2}{2\ell M_2}L\leq\delta n_R<L\,,\,\,\frac{J_2}{\ell M}L\leq\delta n_{tot}<L\,.\nonumber
\end{eqnarray}
In this case, the number of left movers, $n_L$ is decreased due to increase angular momentum, but the total number of states increase. When the particle radial momentum is smaller than $\frac{Jr_h}{2\ell^2M \rho^2}\,$, the changes are,
\begin{eqnarray}
\label{ptccon1}
&&0\leq |p^r|<\frac{J_2r_h}{2\ell^2 M_2 \rho_h}L\,\rightarrow\,-\frac{16G^2\ell^2J_2}{r_h\rho_h^2}L\leq\delta r_h<0\,,\\
&&\frac{4G\ell J_2-\rho_h^2}{2\rho_h^2}L\leq\delta n_L<-\frac{M_2\ell-J_2}{2\ell M_2}L\,,\,\,\frac{4G\ell J_2+\rho_h^2}{2\rho_h^2}L\leq \delta n_R <\frac{M_2\ell+J_2}{2\ell M_2}L\,,\,\,\frac{4G\ell J_2}{\rho_h^2}L\leq\delta n_{tot}<\frac{J_2}{\ell M_2}L\,,\nonumber
\end{eqnarray}
where the horizon size and $n_L$ is decreased, but the entropy increases. Note that the black hole mass and angular momentum are increased due to the particle absorption in above three cases. If the black hole absorbs a particle satisfying eq.~(\ref{ptccon1}), it can slowly attain an angular momentum $J_3\sim \ell M_3$ and bigger mass $M_3$\,. The particle momenta change black hole properties as,
\begin{eqnarray}
0 \leq |p_r|\leq \frac{r_h L}{2\rho_h\ell}\,,\,\,-\frac{4G \ell }{r_h}L\leq\delta r \leq 0\,,\,\,\delta n_R \sim L\,,\,\,\delta n_L \sim 0\,,\,\,\delta n_{tot}\sim L\,,
\end{eqnarray}
where $r_h$ is approximately zero from eq.~(\ref{metric1})\,, such that the decrease in horizon size is almost impossible. When a particle falls into the black hole, its radial momentum will be bigger than zero. The non-zero radial momentum changes the black hole properties such that,
\begin{eqnarray}
|p_r|> \frac{r_h L}{2\rho_h\ell}\,,\,\,\delta r >0\,,\,\,\delta n_R>L\,,\,\,\delta n_L > 0\,,\,\,\delta n_{tot}> L\,.
\end{eqnarray}
Thus, adding mass and rotational energy makes a horizon bigger in size than before in almost all the practical falling. In these cases, the particle energy increases the black hole energy states. The black hole horizon becomes smaller in the specific range of a particle radial momentum, but the range becomes smaller when the particle angular momentum come closer to $M\ell$. Finally, a non-zero radial momentum particle causes to increase the horizon size and the black hole mass, so the horizon does not disappear.



\section{Summary and Discussions}
In a 2+1-dimensional black hole, the black hole thermodynamic properties are investigated by the momenta carried by a particle. The black hole mass and angular momentum are controlled by the particle total energy and angular momentum. This method satisfies the 2nd law of thermodynamics because it increases an irreducible mass that is proportional to the entropy. The rotational energy can be added or extracted by having the particle rotate directions. Therefore, we can avoid unphysical cases that the 2nd law of thermodynamics is not satisfied by the changes in the black hole mass and the angular momentum. Also, particle absorption is the only way in the universe to change black hole mass and rotations. The effects of particle absorption on a black hole evolution are analyzed in specific black hole mass and angular momentum ratios.

Thermodynamic properties are obtained from first-order particle geodesic equations. Using geodesic equations, the particle energy is derived and determined by particle radial and angular momenta. An inequality in eq.~(\ref{bhe1}) gives a relation between black hole mass and particle momenta in particle absorption. In the inequality, a continually increasing property can be obtained, and it is defined as an irreducible mass. The Bekenstein-Hawking entropy is proportional to the square of the irreducible mass. The change in the irreducible mass is only proportional to the particle radial momentum, and so the entropy is also proportional to the particle radial momentum. The particle angular momentum changes the black hole rotational energy. Note that the black hole mass consists of a sum of the irreducible mass and the rotational energy. Using these features, the black hole thermodynamic reactions are categorized into four cases. If a particle has a non-zero radial momentum, the black hole irreducible mass increases, so the entropy also increases. It is a thermodynamically irreversible process. However, the behavior of the black hole horizon and mass are dependent on the relative rotating direction of the black hole and the particle. For the same rotating cases, the black hole mass and rotating energy are larger than they were before absorption. From eq.~(\ref{rh1}), the horizon becomes larger than before if the black hole absorbs a particle with a radial momentum larger than $\frac{r_hJL}{2\ell^2\rho_h M}$\,. For a particle with a radial momentum less than $\frac{r_hJL}{2\ell^2\rho_h M}$\,, the horizon becomes smaller. For the opposite rotating cases, the particle angular momentum decreases the black hole angular momentum, so the black hole rotational energy and angular momentum decrease. The black hole mass is dependent on the balance between the increase of the irreducible mass and the decrease of the rotational energy. If the particle radial momentum is less than $\frac{4GJL}{\rho_h}|L|$\,, the black hole mass increases, and the horizon becomes larger than before, for all radial momentums from eq.~(\ref{rh1}). As a specific case, if a zero-radial momentum particle falls into the black hole, the change in the irreducible mass and entropy is also zero, so it is a reversible process. In this case, the black hole rotational energy only changes in terms of absorption. For the relatively same rotate direction between the particle and the black hole, the angular momentum and rotational energy of the black hole increase. The horizon length always becomes smaller than before in eq.~(\ref{rh1}). For the relatively opposite rotate direction, the black hole rotational energy decreases due to the smaller $M_B$ and $J$ removed particle momenta. The black hole horizon always becomes larger than before in eq.~(\ref{rh1}).

Using AdS/CFT duality, microstate changes are obtained as a CFT interpretation. The entropy change is also rewritten as a left and right quantum number in eq.~(\ref{ent3}), and it is equivalent to what is described as particle momenta. The derived quantum number changes are in eq.~(\ref{deln1}). For a given black hole mass and angular momentum, a particle radial momentum increases $n_L$ as much as $n_R$, but a particle angular momentum provides a positive contribution to $n_R$ and a negative contribution to $n_L$. For increasing a black hole angular momentum, the change of $n_R$ is larger than that of $n_L$, and the total numbers of states $\delta n_R +\delta n_L$ increases. For decreasing a black hole angular momentum, the change of $n_R$ is smaller than that of $n_L$, and the total numbers of states $\delta n_R +\delta n_L$ decreases.

The black hole configuration and microstates are investigated when the black hole increases its angular momentum through a particle absorption from $J=0$ to $J=M\ell$\,. When a $J_1=0$ black hole absorbs a particle that has a radial momentum larger than $L/\ell$\,, the horizon, $n_R$, $n_L$, and $n_{tot}$ are increased. If a particle radial momentum is smaller than $L/\ell$\,, $n_L$ decreases. After the black hole absorbs particles, its angular momentum becomes in the range of $0<J_2<M_2\ell$\,. The effects of a particle are classified in three cases. If a particle has a radial momentum larger than $\frac{\rho_h^2-4GJ_2\ell}{\ell r_h \rho_h}$, the horizon, $n_R$, $n_L$, and $n_{tot}$ are increased. For $\frac{J_2r_h}{2\ell^2 M_2 \rho_h}L\leq |p^r|<-\frac{4GJ_2\ell - \rho_h^2}{\ell r_h \rho_h}L$\,, the left quantum number is only decreased. For $0\leq |p^r|<\frac{J_2r_h}{2\ell^2 M_2 \rho_h}L$\,, the horizon size and $n_L$ can be decreased by adding energy and angular momentum. This 2+1-dimensional black hole angular momentum has an upper bound at $J=M\ell$\,. If the black hole angular momentum increases to $J_3\sim M_3 \ell$\,, the black hole has a very small horizon. In this case, the horizon size, $n_R$, $n_L$, and $n_{tot}$ are increased for a $|p_r|> \frac{r_h L}{2\rho_h\ell}$ particle. However, if a particle has a radial momentum in $0\leq|p_r|\leq \frac{r_h L}{2\rho_h\ell}$\,, the horizon can be smaller than before, in an ideal case. In $J_3\sim M_3 \ell$ case, a horizon closes to zero, but $\rho_h$ is similar to $4GM_3\ell^2$, so making a smaller horizon is only possible in zero-radial momentum cases. When a particle falls into the horizon, the radial momentum is increased by the gravitational force. Thus, a zero-radial momentum particle is practically impossible. The black hole horizon size is increased in 2+1-dimensional spacetime.

{\bf Acknowledgments}

This work was supported by the National Research Foundation of Korea(NRF) grant funded by the Korea government(MEST) through the Center for Quantum Spacetime(CQUeST) of Sogang University with grant number 2005-0049409.

\end{document}